\documentclass{emulateapj}
\usepackage{graphicx}
\usepackage{natbib}
\usepackage{color}
\usepackage{mathtools}

\bibpunct{(}{)}{;}{a}{}{,}

\shorttitle{FRADO model of BLR in AGN}
\shortauthors{Czerny et al.}

\begin{document}


\title{Failed radiatively Accelerated Dusty Outflow model of the Broad Line Region in Active Galactic Nuclei. I. Analytical solution.}


\author{B. Czerny\altaffilmark{1,2}, Yan-Rong Li\altaffilmark{3}, K. Hryniewicz\altaffilmark{2}, S. Panda\altaffilmark{1,2}, C. Wildy\altaffilmark{1}, M. Sniegowska\altaffilmark{1,5}, J.-M. Wang\altaffilmark{3}, J. Sredzinska\altaffilmark{2}, and V. Karas\altaffilmark{4}}
\email{bcz@cft.edu.pl}

\altaffiltext{1}{Center for Theoretical Physics,
    Polish Academy of Sciences, Al. Lotnik\' ow 32/46,
    02-668 Warsaw, Poland}

\altaffiltext{2}{Copernicus Astronomical Center, Polish Academy of Sciences,
    Bartycka 18, 00-716 Warsaw, Poland}

\altaffiltext{3}{Key Laboratory for Particle Astrophysics, Institute of High Energy Physics, Chinese Academy of Sciences, 19B Yuquan Road, Beijing 100049, China}

\altaffiltext{4}{Astronomical Institute, Academy of Sciences, Bocni II 1401, CZ-141 00 Prague, Czech Republic}

\altaffiltext{5}{Warsaw University Observatory, Al. Ujazdowskie 4, 88-478 Warsaw, Poland}

\begin{abstract}
The physical origin of the Broad Line Region in Active Galactic Nuclei is still unclear despite many years of observational studies. The reason is that the region is unresolved and the reverberation mapping results imply complex velocity field. We adopt a theory-motivated approach to identify the principal mechanism responsible for this complex phenomenon.
We consider the possibility that the role of dust is essential.
We assume that the local radiation pressure acting on the dust in the accretion disk atmosphere launches the outflow of material, but higher above the disk the irradiation from the central parts cause the dust evaporation and a subsequent fall back. This failed radiatively accelerated dusty outflow (FRADO) is expected to represent the material forming low ionization lines. In this paper we formulate simple analytical equations describing the cloud motion, including the evaporation phase. The model is fully described just by the basic parameters: black hole mass, accretion rate, black hole spin and the viewing angle. 
We study how the spectral line generic profiles correspond to this dynamics.
We show that the virial factor calculated from our model strongly depends on the black hole mass in case of enhanced dust opacity, and thus it then correlates with the line width. This could explain why the virial factor measured in galaxies with pseudo-bulges differs from that obtained from objects with classical bulges although the trend predicted by the current version of the model is opposite to the observed trend. 
\end{abstract}



\section{Introduction}

The formation of the Broad Line Region (BLR) in active galaxies is not well understood and most likely a number of mechanisms are involved, taking into account the extension of the BLR and large differences in the physical conditions in its various parts. From the point of view of the emitted lines it is generally divided into a Low Ionization Line (LIL) part and a High Ionization Line (HIL) part \citep{collin1986}. In the present paper we concentrate on the LIL part as the lines emitted there, like H$\beta$ and Mg II, are likely to come from the material roughly in Keplerian motion and are considered to constitute a relatively reliable tool for the black hole mass measurement.

The emitted lines are strong so the BLR must be able to intercept a significant fraction of the central radiation and the required covering factor is high. On the other hand, clouds are rarely seen in absorption which hints that the cloud distribution is significantly flattened \citep[e.g.][]{collin2006,gaskell2009}. This may suggest that the BLR is physically related to the underlying accretion disk although this relation is not firmly established. The relation between the disk and the BLR is supported by the disappearance of the BLR at very low luminosities  \citep[e.g.][and the references therein]{czerny2004,balmaverde2015}. On the other hand, successful phenomenological models assume that the clouds are seen both above and below the equatorial plane without the disk blocking the view \citep[e.g.][]{pancoast2011,li2013,pancoast2014}, with the material either inflowing or on elliptical orbits. 

Parametric models can fit the data, but they do not explain why the emitting material behaves in the required way. In this paper we follow a physically motivated approach which is based on a failed dust driven wind from the accretion disk atmosphere.

Accretion disk winds as models of the BLR have been discussed for many years. Those models can be roughly divided into magnetic winds \citep{emmering1992}, line-driven winds  \citep{murray1995} and dust-driven winds. The last option was usually used to mimic the dusty molecular torus \citep{wada2012}. However, dust can exist in the shielded accretion disk atmosphere much closer than in the dusty torus. Such a disk forms a failed wind, since the radiation support due to the dust disappears as soon as the clouds rise high above the disk and become irradiated by the central source. The preliminary version of such a Failed Radiatively Accelerated Dusty Outflow (FRADO) model reproduces the location of the LIL BLR known from reverberation mapping \citep{peterson2004,bentz2009,bentz2013} without any need of an arbitrary constant and predicts the scaling of the BLR size with just the square root of the monochromatic flux, for all values of accretion rates, black hole masses and spins \citep{czhr2011,czerny2015,czerny2016}. 

The advantage of the FRADO model is that it firmly predicts the inner as well as the outer radius of the BLR, with the inner radius set by the accretion disk effective temperature equal to the sublimation temperature, and the outer radius is set by the condition that the dust can survive even if it is fully exposed to irradiation by the central parts. The wind in this model is launched only by the local radiation pressure from the disk since if the material is exposed to the nuclear emission the dust would be destroyed. It is not clear if this local radiation pressure is strong enough to raise the material above the disk. We thus analyse in this paper whether indeed the FRADO model provides a suitable scenario for the dynamics of the LIL BLR. 

\section{The dynamics of the clouds in the FRADO model}

We formulate a simple local model of the dynamics of the BLR clouds assuming a plane-parallel approximation. This allows us to obtain an analytical solution for cloud motion above the accretion disk surface at the given radius, including the effect of radiation pressure acting on the cloud whenever dust is still present, and a subsequent dustless cloud motion after dust evaporation.

The medium close to the accretion disk atmosphere or above it is predominantly in Keplerian motion. However, it might also be the subject of vertical motion if the forces in the vertical direction do not balance. Here we consider only the local situation at a given radius $r$, in the plane-parallel approximation, ignoring the effect of radiation coming from other disk radii. In the absence of a global magnetic field the forces acting on the plasma are the vertical component of the gravity and the radiation pressure from the disk: 

\begin{equation}
{dv \over dt} = - {G M z \over r^3} + {\kappa F_{rad} \over c},
\end{equation}
where $M$ is the black hole mass, $v$ is the plasma velocity perpendicular to the disk, $\kappa$ is the wavelength-averaged effective opacity, $z$ is the coordinate measured from the disk equatorial plane, and $F_{rad}$ is the radiation flux from the disk. Our equations are then different from the 1-D approximation of a radial wind frequently used to describe the later stages of the wind \citep[e.g.][]{dyda2017}. 

The opacity weakly depends on the plasma properties if the source of opacity is electron scattering or dust scattering/absorption. Here we have in mind low temperatures, below the dust sublimation radius, so in this case the dust dominates the opacity. Further we assume, for simplicity, that $\kappa$ is a constant.

The hydrostatic equlibrium at the disk surface ($z = H_{disk}$) requires: 
\begin{equation}
{dv \over dt} = 0, ~~~~~~~~~\Rightarrow~~~~~~~~{G M H_{disk} \over r^3} = {\kappa F_{rad} \over c}.
\end{equation}
If this expression is combined with the expression for the local disk flux in a Shakura-Sunyaev disk \citep{ss1973}, with the inner boundary condition neglected
\begin{equation}
\label{eq:Frad}
F_{rad} = {3 G M \dot M \over 8 \pi r^3},
\end{equation}
it leads to an expression for the disk thickness
\begin{equation}
H_{disk} = {3 \kappa \dot M \over 8 \pi c},
\end{equation}
which is independent from the radius but proportional to the accretion rate. This value is identical to the expression for
the disk thickness in Shakura-Sunyaev disk model if the disk is dominated by radiation pressure and the inner boundary condition is neglected.

Such a simplified analysis provides the hydrostatic equilibrium and does not lead to any wind from the disk. More precise analysis of the dust-supported torus structure in equilibrium has been studied for example by \citet{krolik2007}.

Outflow appears if the description of the disk outer layers is more precise, like in the case of the line-driven outflows \citep[e.g.][]{proga2000}. In the simplest approach we can say that in the disk interior we use the Rosseland mean, $\kappa_R$ while in the disk atmosphere, when calculating the radiation pressure, we should use the Planck mean, $\kappa_P$, where the two quantities are defined in the following way \citep{mihalas1978}:
\begin{equation}
\kappa_P = {\int_0^\infty \kappa_{\nu} B_{\nu} d\nu \over \int_0^\infty B_{\nu} d\nu },
\end{equation}

\begin{equation}
{1 \over \kappa_R} = {\int_0^\infty \kappa_{\nu}^{-1} u_{\nu} d\nu \over \int_0^\infty u_{\nu} d\nu },
\end{equation}
where $B_{\nu}$ is the Planck function and $u_{\nu}$ its derivative with respect to the temperature, $u_{\nu} = \partial B_{\nu} (T)/\partial T$. This approximation for the description of the radiation pressure acting on dusty layer is valid if the dusty shell is optically thick \citep[see e.g.][]{wachter2008}, i.e. the optical thickness of a dusty cloud is above 1. 

The difference between these two quantities drives the outflow, thus at the disk surface we have:
\begin{equation}
\bigl({dv \over dt}\bigr)_{z=H_{disk}} = {(\kappa_P - \kappa_R) F_{rad} \over c},
\end{equation}
and in general:
\begin{equation}
{dv \over dt} = - {G M \zeta \over r^3} + {G M \over r^3}H_{disk}\bigl({\kappa_P \over \kappa_R} - 1\bigr),
\end{equation}
where $\zeta = z - H_{disk}$ is the distance measured from the disk surface. If both $\kappa_P$ and $\kappa_R$ are constant, this equation can be easily integrated analytically to get:
\begin{equation}
\label{eq:v_normal}
v = H_{disk} ({\kappa_P \over \kappa_R} - 1) \Omega_K \sin \Omega_K t,
\end{equation}
and
\begin{equation}
\label{eq:z_normal}
\zeta = H_{disk} ({\kappa_P \over \kappa_R} - 1) \bigl[ 1 -\cos \Omega_K t \bigr],
\end{equation}
where: 
\begin{equation}
\Omega_K = \sqrt{GM \over r^3},
\end{equation}
so the time evolution is described as oscillations between $\zeta = 0$ and $\zeta = 2 H_{disk}(\kappa_P/ \kappa_R - 1)$ with the local Keplerian period. This behavior is very different from the case of stellar winds where the gravity gets weaker as the wind rises up. In the case of a disk geometry this could in principle happen when the coordinate $z$ becomes comparable to $r$. However, here we use the approximation of $z$ much lower than $r$. 

Equations \ref{eq:v_normal} and \ref{eq:z_normal} describe the motion under the condition that the opacity does not change. 
However, high above the disk, the plasma becomes exposed to irradiation by the central source. We ignore the dynamical effect of the additional radial pressure gradient, but we take into account the destruction of the dust grains. 
The phases of motion upward above the disk plane and downward toward the disk become in this case asymmetric.
We now consider the case when the dust evaporates at the time moment $t_0$, when the cloud position $\zeta_0$, and its velocity $v_0$ are set by:
\begin{eqnarray}
\zeta_0 &=& H_{disk} ({\kappa_P \over \kappa_R} - 1) \bigl[ 1 -\cos \Omega_K t_0 \bigr], \nonumber \\
v_0 &=& H_{disk} ({\kappa_P \over \kappa_R} - 1) \Omega_K \sin \Omega_K t_0.
\end{eqnarray}
But now the equations governing the cloud motion change to:
\begin{equation}
{dv \over dt} = - {G M z \over r^3},
\end{equation}
which leads to the analytical solution:
\begin{eqnarray}
\zeta & = & A + B \cos(\Omega t + \phi_0),\nonumber \\
A &=& -H_{disk}, \nonumber \\
B & = & - \lambda H_{disk} {\sin \Omega_K t_0 \over \sin (\Omega_K t_0 + \phi_0) },\nonumber\\
\phi_0 & = & -\arctan {\lambda \sin \Omega t_0 \over \lambda(1 - \cos \Omega t_0) + 1} - \Omega t_0, 
\end{eqnarray}
where $\lambda = \kappa_P/ \kappa_R - 1$.
The highest point reached by the cloud is thus reduced and the amount of reduction depends on the phase of the motion when the dust evaporates.

The inner radius of the BLR within the FRADO model is given by the condition that the effective temperature is equal to the dust sublimation temperature:
\begin{equation}
{3 G M \dot M \over 8 \pi R_{in}^3} = \sigma_B T_{dust}^4,
\end{equation}
so the final expression for the inner radius is
\begin{equation}
R_{in} = \bigl({3 G M \dot M \over 8 \pi \sigma_B T_{dust}^4}\bigr)^{1/3}.
\end{equation}
The height above the disk where the dust evaporates is set by $\zeta_{evap}$, and at $R_{in}$ $\zeta_{evap} = 0$. In general, we thus formulate the condition for the dust evaporation:
\begin{equation}
{L_{central} \zeta_{evap} \over 4 \pi r^3} + F_{rad} = \sigma_B T_{dust}^4,
\label{eq:central_flux}
\end{equation}
where the central luminosity  is given by the accretion rate and the accretion efficiency, $\eta$, as $L_{central} = \eta \dot M c^2$, and the local flux is defined by Eq.~\ref{eq:Frad}. Here we assume that the central radiation comes from a flat accretion disk, thus it is anisotropic, with a $\propto \cos \theta$ dependence on the angle between the symmetry axis and direction towards the cloud, which introduces the factor $\zeta/r$. This gives us an explicit expression for the height where the dust evaporation takes place:
\begin{equation}
\label{eq:zeta}
\zeta_{evap} = {\sigma_B T^4_{dust} 4 \pi r^3 \over \eta \dot M c^2} - {3 \over 2}{1 \over \eta} r_g,
\end{equation} 
where $r_g = GM/c^2$ is the gravitational radius. The expression is of course valid only for $\zeta_{evap} << r$.

\begin{figure}
    \centering
    \includegraphics[width=0.95\hsize]{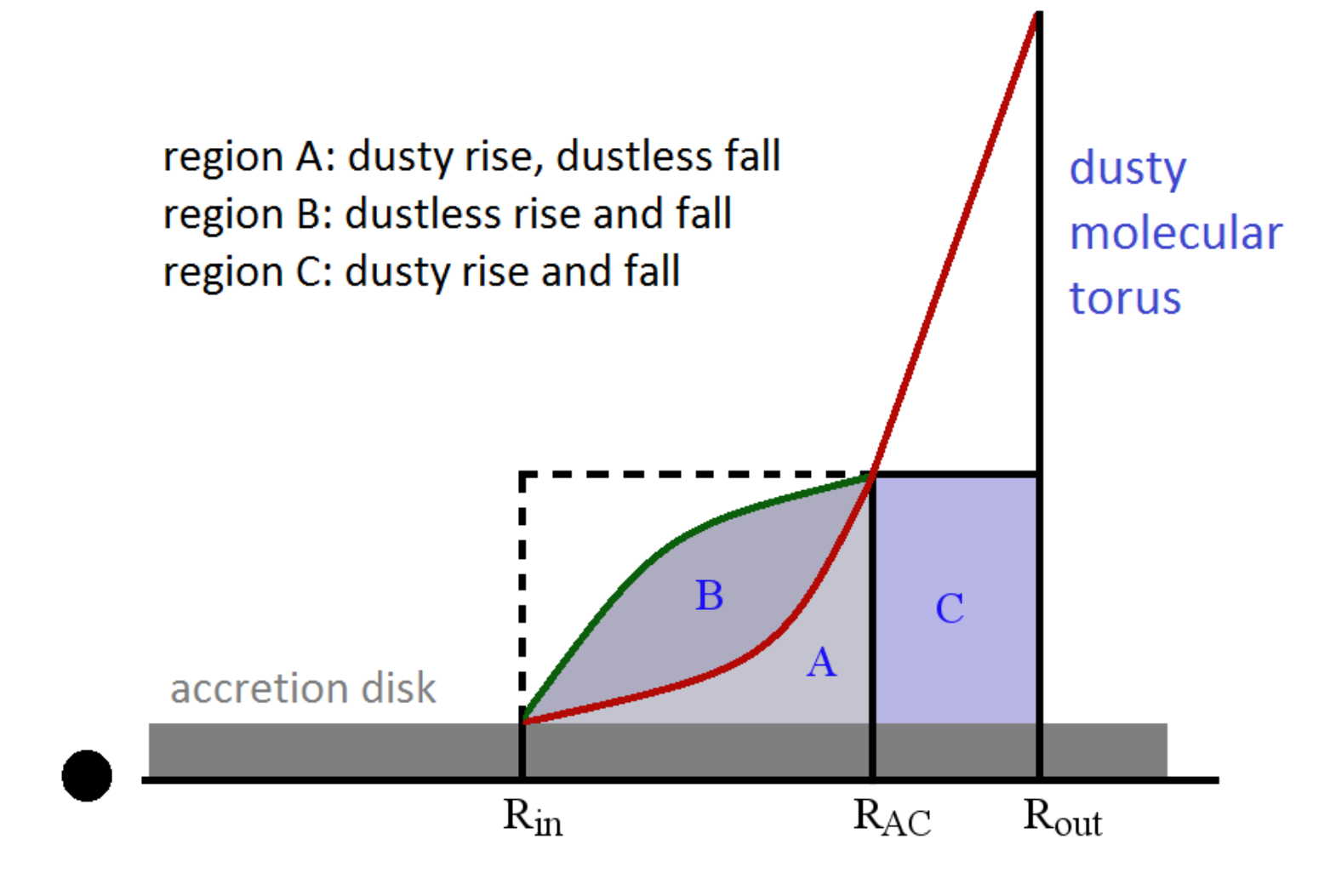}
    \caption{The schematic picture of the BLR with the inner zone populated by a mixture of dusty and dustless clouds, and the outer region populated by dusty clouds. The red line represents the location of dust evaporation.}
    \label{fig:zones}    
\end{figure}

The evaporation height $\zeta_{evap}$ rises rapidly with distance from the black hole. When it is larger than the height of the clouds pushed up by the local radiation pressure there is no evaporation effect. We thus have a complex three
zone structure in the cloud region, illustrated in the Fig.~\ref{fig:zones}. In zone C both the rising and falling clouds contain dust and their up and down velocities are the same. In zone B the clouds are dustless and the up and down velocity is again the same. In zone A the up and down velocities are different, since the rising clouds contain dust and the falling clouds do not contain dust. Thus, overall, the motion is not symmetric.

The outer radius of the BLR or the inner radius of the dusty/molecular torus is set by the condition that $\zeta_{evap} = r$, i.e. the dust survives irradiation by the central parts even if the radiation is spherically symmetric. In general, this is a non-linear equation most easily solved through numerical methods:
\begin{equation}
R_{out} =  {\sigma_B T^4_{dust} 4 \pi R_{out}^3 \over \eta \dot M c^2} - {3 \over 2}{1 \over \eta} r_g,
\end{equation} 
but usually the second term on the right hand side is small so it is possible to use an approximate expression
\begin{equation}
R_{out} \simeq \bigl({\eta \dot M c^2 \over \sigma_B T^4_{dust} 4 \pi} \bigr)^{1/2}.
\end{equation}
Using the last expression, we can derive approximate ratio of the two radii
\begin{equation}
{R_{out} \over R_{in}} \simeq {\eta^{1/2}  c \dot M^{1/6} \over  3^{1/3} (\pi \sigma_B T^4_{dust})^{1/6} G^{1/3} M^{1/3}},
\end{equation}
so the BLR becomes narrower for large black hole masses. The dependence on the accretion rate is weaker, with the rise of the accretion rate whole BLR moves towards larger distances from the black hole.

\section{Emission line profiles}
\label{sect:profiles}

The conversion of the cloud dynamics into motion of individual clouds cannot be done analytically. Still, we keep the scenario as simple as possible, with the aim not to introduce new arbitrary parameters. We introduce a fixed 3-D grid of identical clouds. We use 800 points in the radial direction, and 1500 points in the azimuthal direction. At each ($r$,$\phi$) location a fixed number of 800 clouds is used to sample the vertical motion. We do that uniformly in cloud rise plus fall time.  Next, we need a relative normalization of the emission from individual clouds to the total spectrum. First, we assume that only the central flux contributes to the line production. This is justified since the local disk emission is mostly in the near-IR band. We assume that the central source is anisotropic (see also Eq.~\ref{eq:central_flux}), it is located at height equal to the disk thickness, $H_{disk}$, and the incident flux thus scales with cloud height as $(z-H_{disk})/r$. We assume that all clouds are identical, so the amount of central flux available to the clouds at a given radius scales as $dr d\phi/r^2 $, with the factor $1/r^2$ accounting for the dilution of the central flux. This again seems justified if the total optical depth of the cloud distribution is constant as a function of radius. Combining this with the central flux anisotropy we assume the cloud emissivity in the form of $(z-H_{disk}) dr d\phi/r^3$. We do not discuss the proportionality constant which, in principle, would provide the covering factor since the model is still in early test phase. Second, we do not take into account that clouds in the outer part of the BLR can be shielded by the innermost clouds. This likely leads to over-estimation of the role of the outer part of the BLR and may cause lines to appear narrower. However, the proper computations of the single cloud contribution with the self-shielding included would require first an ad hoc assumption of the cloud size as a function of radius and vertical height, and then complex modelling of the shielding. Thus the advantage of our modeling is the simplicity, with reasonable scaling of the different BLR clouds to the total emission  but the disadvantage is that we do not predict the covering factor of the BLR as a function of the global parameters. We will have to address this issue in the future. 

Clouds are then viewed by an observer at the inclination angle $i$ with respect to the symmetry axis. Since we consider clouds which are optically thick, we take into consideration the Moon-type effect as for example in \citet{goad2012}, so only a part of the cloud facing the central source is illuminated by the central source and emits the radiation. This gives an additional flux weight to a single cloud emissivity located at (r,z):
\begin{eqnarray}
\cos \gamma & = &{(z_S-z)\cos(i)  - y \sin(i) \over (r^2 + (z-z_S)^2)^{1/2}},\nonumber \\
f_{Moon} & = & 1 - \gamma/\pi,
\end{eqnarray}
where $z_S$ is the central source height (in our model we assume $z_S = H_{disk}$).

In the present model we assume that a constant fraction of the incident radiation from the central source is changed into line emission. In principle we should calculate the line emissivity using, for example, the code {\sc cloudy}, which allows us to take into account the effect of dust on the line formation efficiency \citep[see e.g.][and the references therein]{adhikari2016}. However, this would require specification of the local density and column density as a function of cloud position which would lead to a multi-parameter complex model. We thus develop a simple universal model for all lines, neglecting the effect of the actual line formation efficiency. On the other hand, we describe the dynamics of the clouds consistently with the FRADO model, and we concentrate on the study of the line profiles which result from such a picture. The local emission is set as a delta function, and the observed line wavelength from a given cloud is given by its radial velocity with respect to the observer. 
We employ an approximation to Doppler (special relativity) shift and the 
gravitational (general relativity) shift of the spectral-line energy, 
i.e., the change of observed vs. intrinsically emitted (rest-frame) 
wavelength, $g \equiv E_{\rm obs}/E_{\rm em}$. This allows us to capture 
the effect of line splitting into two horns due to fast, predominantly 
orbital motion of the emitting material. We do not include the complete 
treatment of these effects, which become important only very close to 
the black hole, at a distance of tens of gravitational 
radii. In particular, the adopted approximation neglects the 
relativistic change of the radiation intensity in the line (factor 
$g^3$; \citealt{laor1991}), which means that the height of the red wing of 
an emission line is the same as the height of the blue wing. Whereas the 
radial component of the motion causes the broadening and the split of 
the observed line shape, the transverse Doppler shift and the 
gravitational shift produce an overall redshift of the line towards 
lower energy ($g<1$; \citealt{karas1992}). Also, the effect of light 
bending contributes to the variation of the observed flux if the source 
orbital plane is at high inclination with respect to the observer. 
However, the latter effects become negligible at distance above several 
tens $r_g$, so we do not need to take them into account within the level 
of accuracy and sophistication of our basic model.

In our picture only clouds above the disk contribute to the line emission; clouds forming at the other disk side are shielded by the disk from the observer.

The model parameters are: the black hole mass, $M$, the accretion rate, $\dot M$, accretion efficiency, $\eta$ (or spin), and the viewing angle, $i$. For convenience, we frequently use the dimensionless accretion rate, $\dot m$, defined as $\dot M /\dot M_{Edd}$, where $L_{Edd} = \eta \dot M_{Edd} c^2$. The additional physical parameters are: the dust sublimation temperature, $T_{dust}$, and the Planck mean opacity, $\kappa_P$. We set the last two quantities at 1500 K and 8 cm$^2$g$^{-1}$ in most models but we check the sensitivity of the results to those two parameters as well. We thus neglect the gas opacity which is small ($\sim 2 \times 10^{-4}$ cm$^{2}$g$^{-1}$, \citealt{bowen1988}, see also \citealt{schirrmacher2003}). The Rosseland and Planck mean were discussed by \citet{semenov2003}, and our adopted value is somewhat larger than recommended close to the sublimation temperature. We discuss this issue later on. 

\section{Results}

\begin{figure}
    \centering
    \includegraphics[width=0.95\hsize]{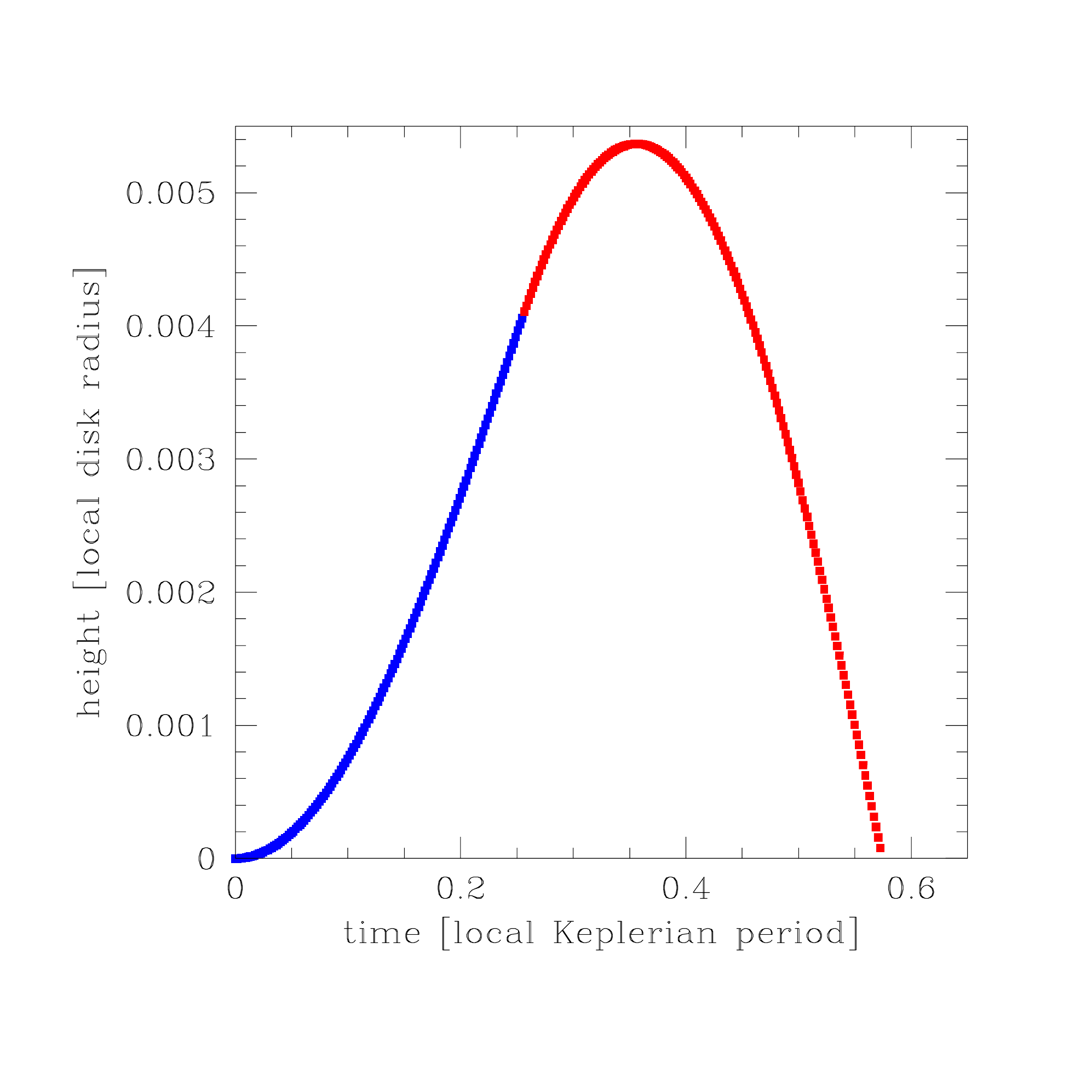}
    \caption{The example of the cloud rise and fall with a clear asymmetry between the two due to the dust evaporation followed by a ballistic motion. A dusty cloud - blue points, a dustless cloud - red points. Parameters: $M = 10^7 M_{\odot}$, $\dot m = 1.0$, $r = 8.5 \times 10^{16}$ cm. The corresponding Keplerian period is 135 yr.}
    \label{fig:motion}    
\end{figure}

The FRADO model predicts the motion of the clouds within the BLR. This dust-driven motion, due to dust evaporation at high altitudes, shows clear asymmetry between the rise and fall. The exemplary plot of the elevation of a single cloud as a function of time is shown in Fig.~\ref{fig:motion}. In particular, the rise starts with zero velocity but the cloud falls back with non-zero velocity, hitting the disk surface. Since the cloud spends more time on the rising part while the fall back is more rapid, the net effect would imitate an outflow, although the material does not leave the region. 

\subsection{Emission line profiles in the standard FRADO model}

\begin{figure*}
    \centering
    \includegraphics[width=0.95\hsize]{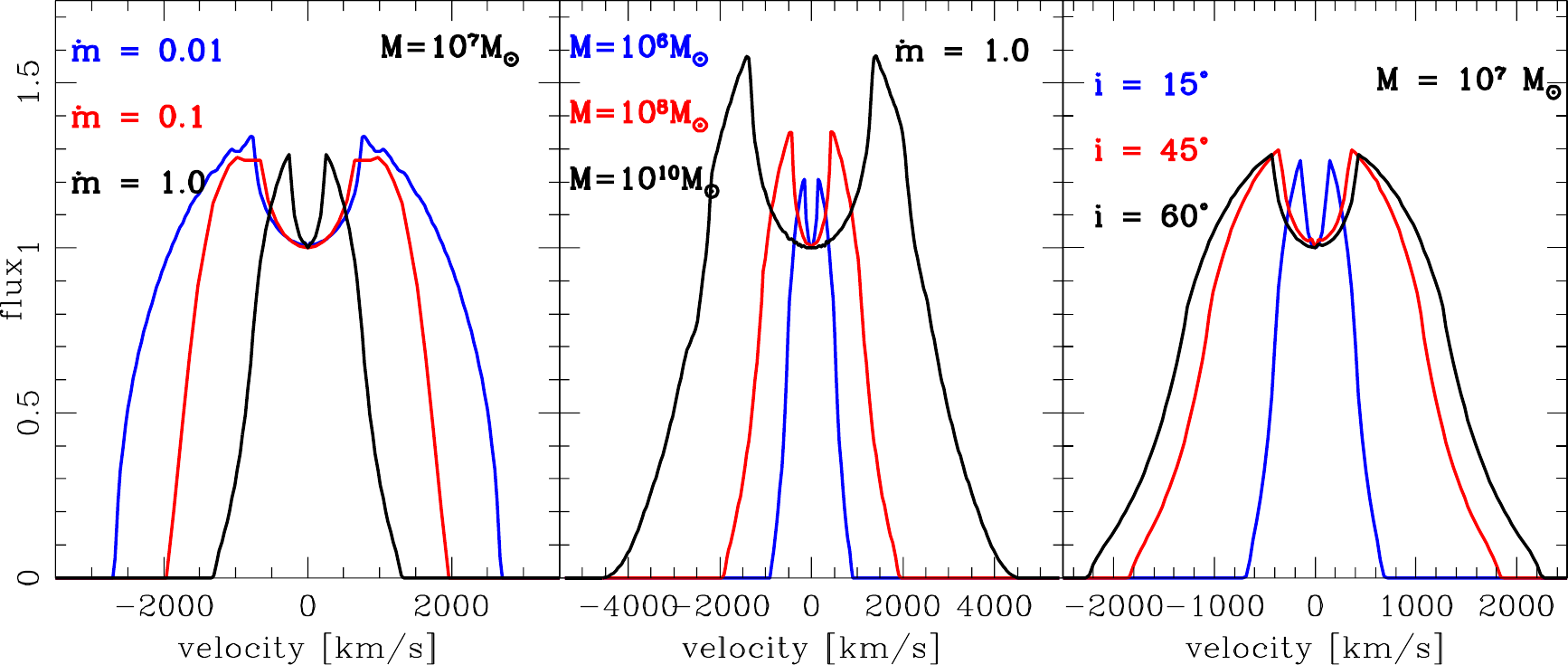}
    \caption{The predicted dependence of the line profile on the accretion rate, the black hole mass and the inclination angle. Unless specified, the parameters are: black hole mass $10^7 M_{\odot}$, accretion rate $1.0$, and viewing angle $30^\circ$.}
    \label{fig:profiles}    
\end{figure*}

We now calculate the line profiles for a grid of models. We fix the accretion efficiency $\eta$ at 0.1, the dust temperature $T_{dust}$ at 1500 K, as appropriate for carbon grains \citep{sharp1995,duschl1996,henning2013}, and adopt $\kappa_R = 2$ cm$^2$g$^{-1}$, $\kappa_P = 8$ cm$^2$g$^{-1}$ \citep{semenov2003}. The grid covers a range of masses from $10^6 M_{\odot}$ to  $10^{10} M_{\odot}$ in logarithmic step. The considered range of  accretion rates is appropriate for bright AGN, with the dimensionless accretion rate between 0.01 and 1.0, also logarithmically spaced. We do not consider accretion rates lower than 0.01 since in case of low luminosity AGN the inner Advection-Dominated Accretion Flow (ADAF) can remove the part of the innermost BLR \citep[see e.g.][]{czerny2004}, which would require a specific adjustment of the model. For the viewing angles, we take values from 5 to 60 deg, representative for type 1 AGN not obscured by the dusty molecular torus. The left panel in Fig.~\ref{fig:profiles} presents the influence of the accretion rate, the middle panel shows the effect of the black hole mass change, and the right panel illustrates the dependence on the inclination angle. 
   
The lines are clearly narrower when the Eddington ratio is higher, in concordance with the generally accepted trend. Nearby type 1 AGN are divided into Narrow Line Seyfert 1 (NLS1) galaxies and Seyfert 1 galaxies, as introduced by \citet{osterbrock1985}. NLS1 have narrower lines and they are also higher Eddington ratio sources \citep{boroson1992,pounds1995,grupe2004,du2016}. The dependence on the black hole mass is also consistent with the observed trends. Seyfert galaxies are divided into NLS1 and BL S1 at the Full Width Half Maximum (${\rm FWHM}$) of 2000 km s$^{-1}$, but the corresponding division in quasars into type A and type B is done at the higher line width, 4000 km s$^{-1}$ \citep{sulentic2000}. Typical black hole masses of Seyfert galaxies are $\sim 10^7 M_{\odot}$ while quasar masses are closer to $10^9 M_{\odot}$. According to our model, an increase in the black hole mass by a factor of 100 leads to an increase of the ${\rm FWHM}$ by a factor $ 2.0$ for $\dot m = 1.0$. This change does not simply result just from the shift of the BLR but reflects the overall change in the region: its relative extension ($R_{out}/R_{in}$) drops from 30 down to 14 with an increase of the black hole mass. 

However, there are two major problems in our basic model: the predicted lines are too dominated by the disk-like shape and the rise of the material above the disk surface is not high enough, even in the sources at the high Eddington ratio. 

The first problem is most likely related to the simplified treatment of the emissivity. Wind models having elements of the radiative transfer show single-peaked lines \citep{murray1995,waters2016}. The second problem may be more difficult to reconcile with our present scenario.

In the case of $10^7 M_{\odot}$, the vertical coordinate saturates at $6.7 \times 10^{14}$ cm, and the maximum $z/r$ ratio is only about 0.006. The vertical velocity is also a small fraction of the local Keplerian speed. In the case of larger black holes this ratio is higher, at 0.05, but still small. The radiation force acting on dusty clouds that comes from the disk below is not strong enough, unless the accretion disk is highly super-Eddington, and we already noticed this difficulty previously, even assuming a very simple model \citep{czerny2015}. If we calculate a highly Eddington model, for a large black hole mass, then the gas reaches higher altitudes, but the lines become narrower as the BLR moves out. The line becomes asymmetric as in this case the vertical velocities are almost comparable to the Keplerian velocity and the motion asymmetry illustrated in Fig.~\ref{fig:motion} becomes important.  

\subsection{${\rm FWHM}$/$\sigma$ trends}

\begin{figure*}
    \centering
    \includegraphics[width=0.95\hsize]{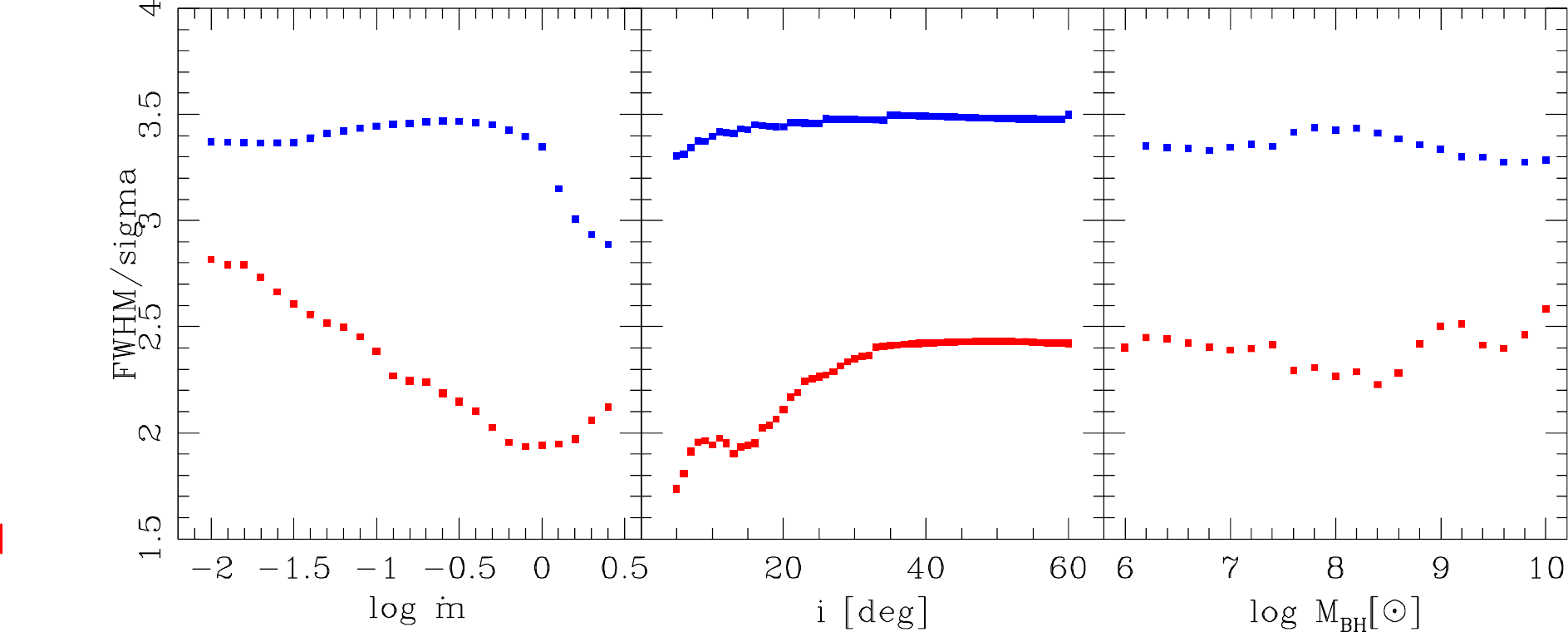}
    \caption{The dependence of the ratio of ${\rm FWHM}$ to $\sigma$ on the accretion rate, the viewing angle and the black hole mass for $\kappa_P = 8$ (the blue points) and  $\kappa_P = 800$ (the red points). For the reference, we choose a black hole of mass $3 \times 10^7 M_{\odot}$, accretion rate $\dot m = 0.1$, and viewing angle $30^{\circ}$.}
    \label{fig:FWHM}    
\end{figure*}

The important parameter of the line shape is the ratio of the ${\rm FWHM}$ to the dispersion $\sigma$ \citep{collin2006}. We thus calculated a sequence of models illustrating the dependence of this ratio on the accretion rate, viewing angle and the black hole mass for two cases of the Planck opacity. The results are plotted in Fig.~\ref{fig:FWHM}. We do not predict a strong dependence of that ratio on accretion rate, unless the accretion rate approaches the Eddington ratio, or inclination, apart from the smallest inclinations below 20 deg.  There is a weak dependence of the line shape on the black hole mass despite systematic trend in the BLR extension in our model. If we fix the dimensionless accretion rate, the ratio of the outer to the inner radius varies with mass. For example, for $\dot m = 0.6$, $R_{out}/R_{in} = 100 $ for $3 \times 10^5 M_{\odot}$ and 20 for  $3 \times 10^6 M_{\odot}$. The range of accretion rates used in the computations is smaller, but also the dependence of the BLR extension is weaker than given by Eq.~10 of \citet{czhr2011} since we now use the condition of $\zeta_{evap} = r$ in Eq.~\ref{eq:zeta} to get the BLR outer radius instead of a simple condition based just on the bolometric luminosity. This weak dependence on BLR extension is caused by the fact that for low black hole masses the contribution of the outermost part of the BLR to the line flux is lower than in the case of large black hole mass, which compensates shrinking full size of the BLR with black hole mass. 

The mean value is higher than the typical values in the observed samples. For example, the sample in \citet{peterson2004} has the ratio $2.03 \pm 0.59$. This is because the lines are slightly too narrow in the basic version of the model.

\subsection{Virial factor}

The understanding of the motion of the BLR clouds is essential for black hole mass measurement.  Classical description of a spherically symmetric cloud distribution in purely Keplerian motion gave an analytic value of the virial factor of $\sqrt{3}/2$ \citep[e.g.][]{kaspi2000}, but flattening of the BLR and likely outflows imply some departure from this simple picture. Therefore, usually the virial factor is based on some external scaling. However, models of the BLR dynamics can provide those factors. As shown by \citet{kashi2013}, magneto-centrifugal winds can lead to considerable departures from rotationally supported cloud motion while radiation pressure driven winds preserve the angular momentum, accelerate slowly, and are roughly virialized. Specific wind models were also discussed by \citet{yong2016}, and gave constraints consistent with the results of \citet{pancoast2014} obtained from modeling the reverberation measured sources.

Our model does not contain any free parameters, if the dust sublimation temperature and opacities are fixed. The FRADO model contains information about the inner radius of the BLR, the black hole mass and the ${\rm FWHM}$ of the line for a given viewing angle. This allows us to calculate the virial factor from the model:
\begin{equation}
f_{vir} = {G M \over {\rm FWHM}^2 R_{in}}.
\end{equation}
Obtained values are slightly larger than generally adopted \citep[see e.g.][and the references therein]{dupu2017},  since there is not much emission very close to the $R_{in}$ region itself. The current model is purely stationary, and we do not yet predict the effective radius, which would be measured by observing the delay of the line. The best current estimate based on the $M - \sigma_*$ relation is $f_{\sigma} = 4.82 \pm 1.67$ \citep{batiste2017} but values of  $f_{FWHM}$ are roughly by a factor of 4 smaller due to typically factor of 2 difference between FWHM and $\sigma$ in line measurement.
The model does not predict strong dependence of the virial factor on the black hole mass or accretion rate since the departure from the Keplerian motion is not strong. However, the viewing angle is clearly important because of the mostly Keplerian character of the motion.  

\begin{figure}
    \centering
    \includegraphics[width=0.95\hsize]{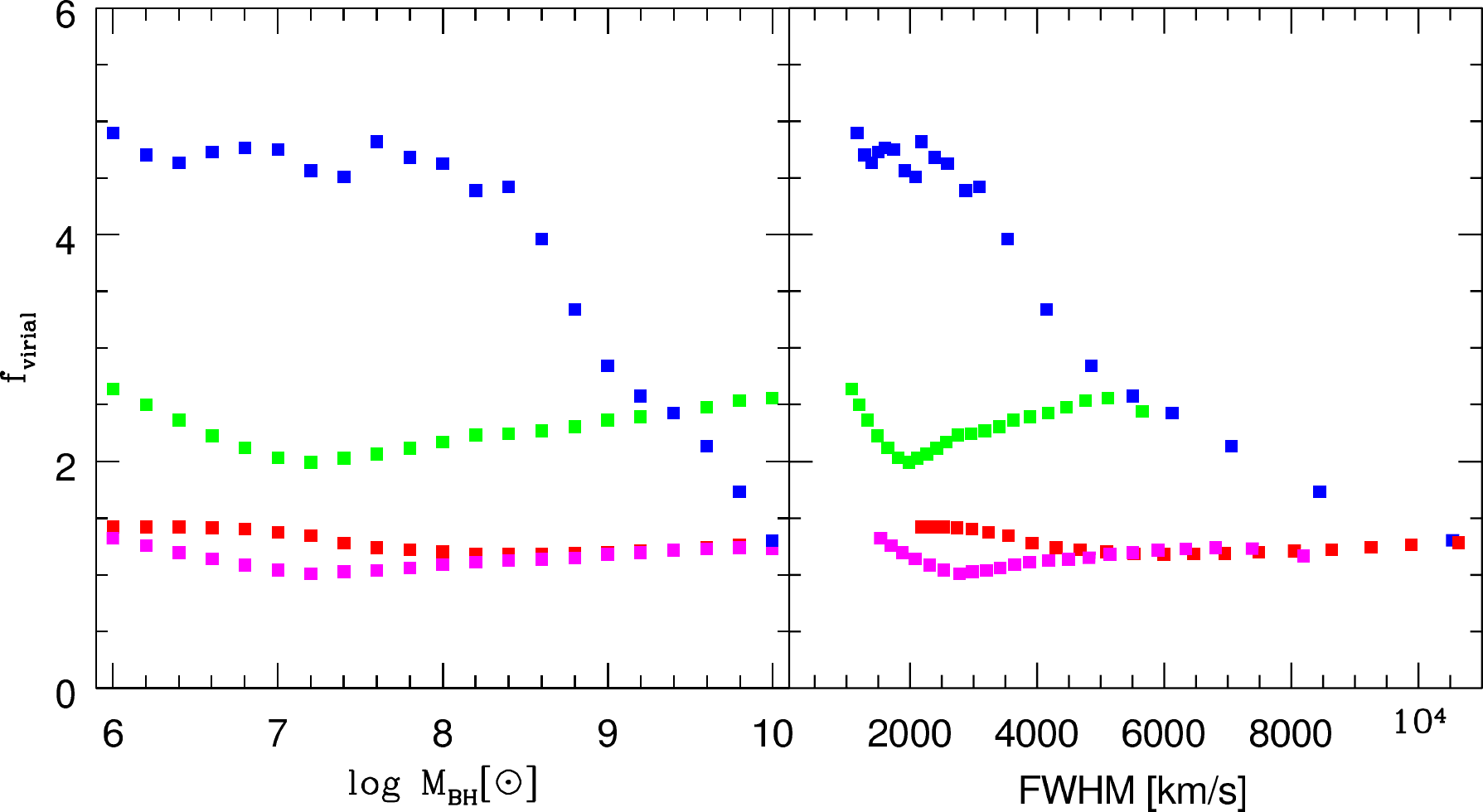}
    \caption{The dependence of the virial factor predicted by the FRADO model on the black hole mass (left panel) and FWHM (right panel) for $i = 30^{\circ}$ and $\dot m = 0.1$ (red points), $i = 30^{\circ}$ and  $\dot m = 1.0$ (green points), for $i = 45^{\circ}$, $\dot m = 1.0$ (magenta points), and again for $i = 30^{\circ}$ and $\dot m = 0.1$ but for enhanced opacity $\kappa_P = 800$ cm$^2$ g$^{-1}$ (blue points).}
    \label{fig:factor}    
\end{figure}

From the observational point of view it is more convenient to express this dependence of the virial factor on the measured line parameter, ${\rm FWHM}$. We plot the result in the right panel of Fig.~\ref{fig:factor}. Again, the dependence on the black hole mass and accretion rate is weak but of course the inclination angle plays an important role.

\begin{figure*}
    \centering
    \includegraphics[width=0.95\hsize]{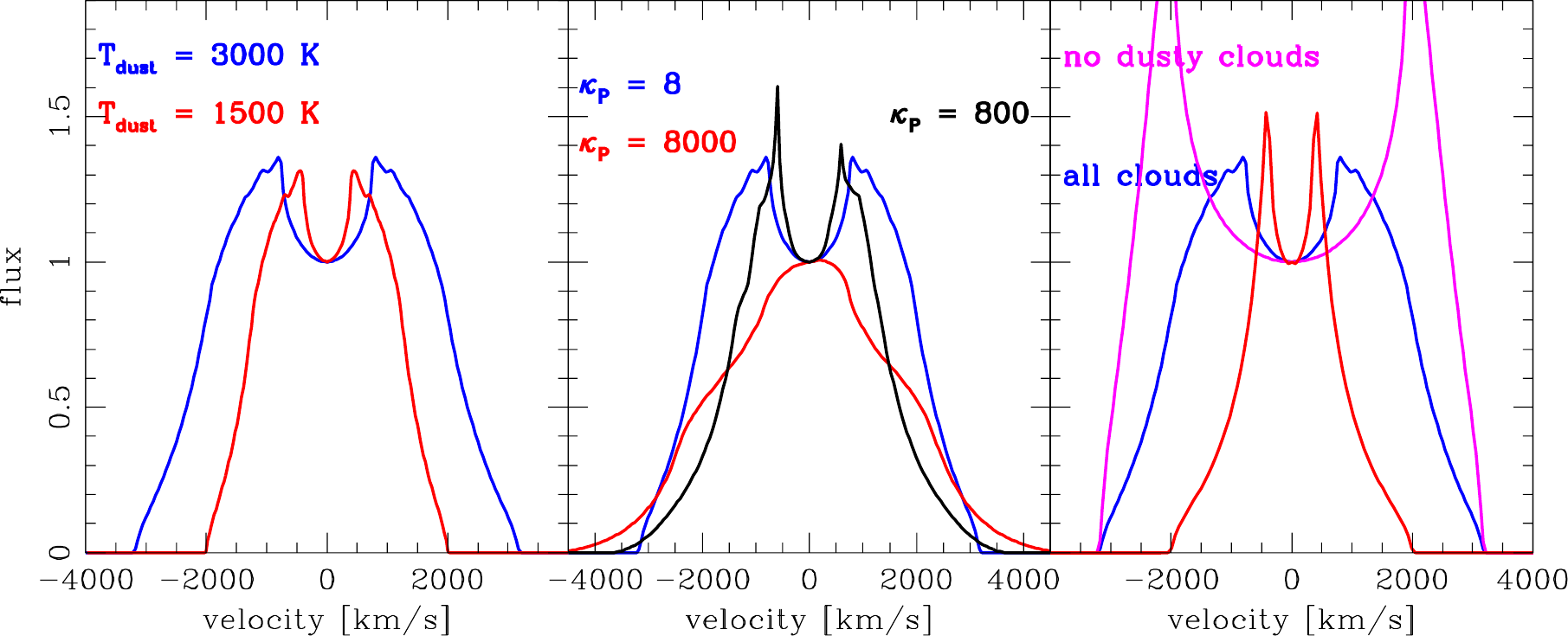}
    \caption{The sensitivity of the predicted line profiles on the model assumptions: the dust sublimation temperature (the left panel), the Planck opacity (in cm$^2$g$^{-1}$, the middle panel) and dust not affecting the line emissivity (the right panel). The parameters of the reference model are: the black hole mass $10^7 M_{\odot}$, the accretion rate $\dot m = 0.6$, the inclination angle $45^{\circ}$, $T_{dust} = 3000$ K. Additionally, red curve in the right panel illustrates the change of irradiating flux geometry from disk-like to isotropic.}
    \label{fig:rozne}    
\end{figure*} 

\subsection{Line dependence on the adopted model parameters}

The lines resulting from our basic models are always double-peaked. Double-peaked profiles are a characteristic signature of the line
formation from a planar distribution of the line-emitting material in orbital motion \citep[e.g.,][]{eracleous1994,karas1992}, and low Eddington ratio sources show such a structure in their lines.

However, observations show that high Eddington ratio sources have rather single-peaked lines, which is not reproduced in the basic version of our model. Also the line kinematic width is generally too narrow in our models, FWHM reaches 7 000 km s$^{-1}$ for black hole mass $10^7 M_{\odot}$, accretion rate $\dot m = 0.01$ and inclination $45$ deg, and only in the case of the highest black hole masses in excess of $10^9 M_{\odot}$ the line width approaches 10 000 km s$^{-1}$ for a representative accretion rate for distant quasars $\dot m = 0.1$ and the same viewing angle.
Also the predicted covering factor measured by the predicted maximum vertical extension of the cloud distribution is certainly too small as indicated by Fig.~\ref{fig:motion}. Our model formally does not have free parameters, but it is nevertheless based on some numerical values adopted in the model. We thus study the sensitivity of the line profiles on the basic model assumptions.

\subsubsection{Dust sublimation temperature}

The dust sublimation temperature is strongly dependent on the dust chemical composition and the dust particle size. The value adopted in our basic model seems to be representative \citep{kobayashi2011,mor2012}. On the other hand, stars forming the dust efficiently in their winds generally have higher effective temperatures than the dust sublimation temperature as the dust actually forms at a larger radius than the location of the stellar photosphere. The effective temperatures of AGB stars forming dust set around 2800 K \citep[e.g.][]{freytag2017}. The dust formation mechanism in those stars requires stellar pulsations and time-dependent temperature above the photosphere. This condition might be fullfilled in the case of AGN disks since accretion disk instabilities (ionization instability, e.g. \citealt{janiuk2011}, and self-gravitational instability, e.g. \citealt{wang2011}) may lead to local disk variability. Thus we decided to vary the dust evaporation temperature over a physically reasonable range in our model
computations.
In Fig.~\ref{fig:rozne},left panel, we show an example of the line profile for black hole mass $10^7 M_{\odot}$, $\dot m = 0.6$ and $i = 45^{\circ}$, for two values of the dust sublimation temperature: 1500 K (standard value) and 3000 K. The line is clearly broader (in our example ${\rm FWHM}$ increased by a factor of 1.8), but the double-peaked structure is more visible even for an Eddington ratio of 0.6. Thus the increase in the dust sublimation temperature does not fully solve the problem. However, it is important to notice that in the FRADO model the determination of the black hole mass based on the measurement of the ${\rm FWHM}$ would depend on the precise value of the maximum dust temperature. 

\subsubsection{Planck opacity and cloud optical depth}
\label{sect:Planck}

The dust opacity used in our basic model is determined highly inaccurately. First, the dust properties depend on the dust composition and dust grain sizes. Secondly, if the dust-gas coupling is not strong enough, some separation into dusty and dustless material can take place, with the dusty material being then the subject of much stronger acceleration. Thirdly, if the dusty material is optically thick, then not just the radiation momentum, but also the radiation energy can be used for the acceleration of the material, which would mimic much larger dust opacity. Therefore, we explore the dependence on the dust opacity is a very broad parameter range. 

Two examples of the line profile obtained from such an approach are shown in the middle panel of Fig.~\ref{fig:rozne}. For $\kappa_P = 8000$ cm$^2$g$^{-1}$, the height of the BLR clouds is comparable to the radial distance. This allows them to intercept a significant fraction of the photons from the central region and produce a strong emission line. The line becomes single-peaked like typical quasar lines, although the line is still not very broad. Further increase of $\kappa_P$ leads to a vertical cloud speed much larger than the Keplerian speed and the height $z$ much larger than $r$. The clouds in our solution would still fall back because of the approximation to the gravitational potential we use. Thus this kind of solution is clearly unphysical, as in this case clouds would form a wind, instead of a failed wind, therefore the equations of motion must be modified and our analytical solution cannot be used.  

\begin{figure}
    \centering
    \includegraphics[width=0.95\hsize]{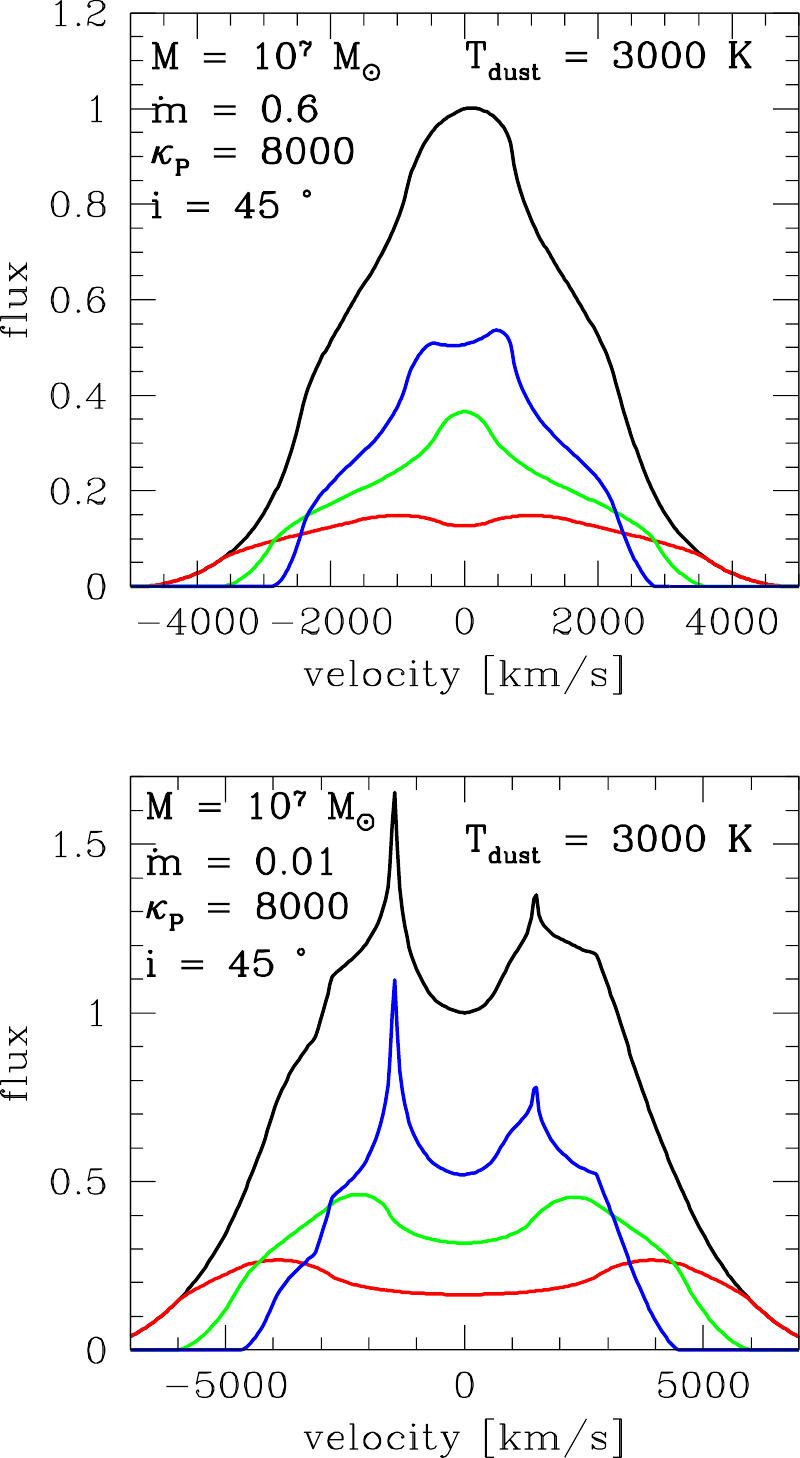}
    \caption{Two examples of a line decomposition into the contributions from the inner BLR (the red line), the middle BLR (the green line) and the outer BLR (the blue line), which sums up to the overall line profile (the black line). The model parameters are: $M = 10^7 M_{\odot}$, $\kappa_P = 8000$ cm$^{2}$g$^{-1}$, $i = 30^{\circ}$, $T_{dust} = 3000$ K for both cases, $\dot m = 0.6$ (upper panel) and $\dot m = 0.01$ (lower panel). If only the inner part of the BLR is considered, the lines are much broader than the whole profile indicates.}
    \label{fig:decompo}    
\end{figure}

In Fig.~\ref{fig:decompo} we show the decomposition of the line profile in the case of an increased $\kappa_P$ into contributions from the inner, middle and outer parts of the BLR, chosen as logarithmic steps in radius. The overall line shape lost its double-peak character due to the large vertical velocities in the inner and middle parts of the BLR while still preserving the double-peak character in the outermost region where velocities are small even for a such high value of the Planck mean. The overall line shape resembles more the typical lines in Narrow Line Seyfert 1 galaxies. However, the overall shape is better represented by a single Gaussian than by a single Lorentzian, the wings of the line not being broad enough to give the Lorentzian shape.  The picture also shows that the emission comes in equal proportions from all three zones, so the flux-weighted emission comes not from the inner part of the BLR, but from the logarithmic mean between the inner and the outer radius. This is why the best fit of the dust temperature to the BLR distance measured from the reverberation is only 900 K instead of 1500 K, or more. This also explains why the ratio of the outer to inner radius of the BLR in our model is much higher (factor 30 or more) than the measured ratio between the BLR size and the dusty-molecular torus size \citep[about 5,][]{koshida2014}, and the virial factor calculated on the basis of $R_{in}$ is too high. This figure also illustrates the importance of the proper accounting for the contribution from different parts of the BLR. If the contribution of the outermost region is neglected, the FWHM of the line gets as high as 3000 km s$^{-1}$ for high accretion rate case and 6000 km s$^{-1}$ for low accretion rate case for the adopted mass of $10^7 M_{\odot}$. Lines are even broader for larger black hole masses. We thus see that to explain well the covering factor and the line width in the future we need more careful modeling of the radiation pressure as well as of the relative contribution of the different parts of the BLR to the final profile.

The enhancement of the dust opacity increases the role of the vertical velocity strongly affects the line shape and the value of the predicted virial factor. The line wings are now better developed, and the typical values of the FWHM/$\sigma$ become comparable to the observed values (see Fig.~\ref{fig:factor}). Now
 the model predicts the dependence of the factor on the black hole mass (see Fig.~\ref{fig:factor}, left panel). Therefore, using the same factor for Seyfert galaxies and massive quasars may lead to a systematic error by a factor of 2. 

The scaling of the virial factor with the black hole mass and the measured ${\rm FWHM}$ has very interesting consequences. It may solve the puzzling result regarding the dependence of the measured virial factor on the type of the host galaxy. \cite{hokim2014} showed that the virial factor measured in galaxies with bulges and with pseudo-bulges differ by a factor of $\sim 2$. From their Fig.~1 we see that those systems actually differ with respect to the mean value of the ${\rm FWHM}$, AGN in pseudo-bulges have on average narrower emission lines and smaller masses. Our correlation resulting from the FRADO model shows that some correlation of the virial factor with the bulge type is expected. However, the trend predicted by the analytical version of FRADO model goes in the opposite direction: virial factor is expected to be lower, not higher, for galaxies with pseudo-bulges. It remains to be seen whether the improved, numerical version of the model would reverse this trend.
Also the values of the virial factor now become too large because of larger contributions from outer parts of the BLR, which is against the required trend. It indicates that the increase in the dust opacity has to be combined with some other effects to obtain the agreement with the observed line properties.

\subsubsection{Dust suppression of the line emission}

It is generally accepted that the presence of the dust suppresses the line emissivity, as demonstrated  in a seminal paper by \citet{laornetzer1993}, where they showed that the separation between the BLR and the NLR is caused by the dust intercepting ionizing photons. On the other hand, this effect strongly depends on the dust properties, the local density of the cloud and on the shape of the incident continuum, so the effect is not always present \citep[see e.g.][]{adhikari2016}. In our basic model we considered the line emissivity as independent from the presence or absence of the dust. In this section we check whether the line profile is significantly modified when the dusty clouds do not emit the line. The result is shown in Fig.~\ref{fig:rozne}, right panel. The removal of the emission from the dusty clouds leads to a broader but disk-like shape for typical values of the model parameters. This trend does not bring the model closer to the observed line shapes. The line asymmetry is not well visible, since the Keplerian velocity dominates.

\subsubsection{Isotropic central source}

The cloud dynamics has been obtained under the assumption that the central flux comes from a flat inner disk and shows the linear dependence on the cosine of the inclination angle. However, in the classical wind papers spherically symmetric central flux has been adopted \citep[e.g. ][]{murray1995}. To test the sensitivity of the line profile on the adopted symmetry we calculated exemplary line shape under assumption of spherically symmetric central flux, without $z/r$ factor included. This example is shown in Fig.~\ref{fig:rozne}, right panel. The line has now better developed wings since part of the emission comes from the innermost region of the BLR, but the overall FWHM is narrower for the same values of all the other parameters, since the emissivity is then more spread across the whole BLR. For isotropic central source, with no shielding of outer clouds by inner clouds and under assumption of equal cloud number at each radius the line emissivity is evenly spread due to the drop of the central flux with radius as $r^{-2}$ and an increase in the emitting area as $r^2$ from logarithmic bins in radius. This type of emissivity also leads to strong dependence of the virial factor on black hole mass even for standard value of the dust opacity since emissivity is less concentrated around $R_{in}$.

\section{Exemplary application of the FRADO model to quasar CTS C30.10}

In order to assess more quantitatively the shortcomings of the present simple analytical model, we compare the model to an exemplary emission line shape.

\begin{figure}
    \centering
    \includegraphics[width=0.95\hsize]{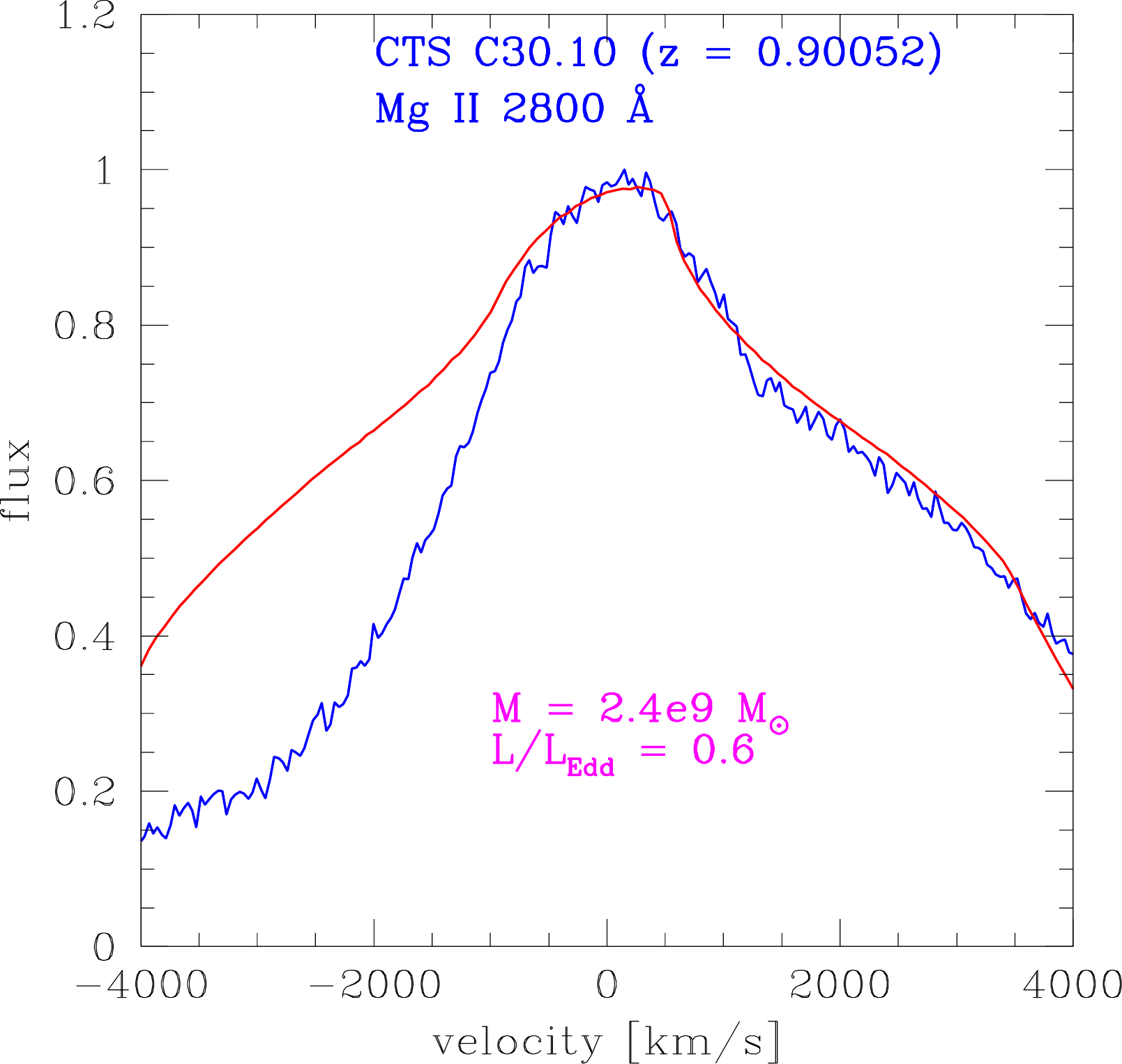}
    \caption{The comparison of the FRADO model prediction (red smooth curve) with the shape of the Mg II line in the quasar CTS C30.10 (blue curve) after we allowed for an artificially increased dust opacity $\kappa_P = 1200$ cm$^2$ g$^{-1}$. The model reproduces well the line core, including the second blue component characteristic for type B quasars, but it does not predict the red Lorentzian tail.}
    \label{fig:CTS}    
\end{figure}

The Quasar CTS C30.10 has been monitored by the Southern African Large Telescope (SALT) since 2012 in the spectral band of Mg II $\lambda2800$ \AA. The shape of the spectrum does not seem to change \citep{modzelewska2014}, and the amplitude changes are moderate (Sredzinska et al., in preparation). We thus combined 17 spectra of that quasar collected in the period from December 2012 to December 2016 to obtain a single average spectrum. The mean quasar magnitude in the V band is 17.067, so the  monochromatic flux at 3000 \AA~ in the quasar rest frame is 46.083 \citep{kozlowski2015} for the adopted source redshift $z = 0.90052$ and the cosmological parameters $H_0 = 70$ km s$^{-1}$ Mpc$^{-1}$, $\Omega_{mat} = 0.28$ and $\Omega_{\Lambda} = 0.72$. For the black hole mass and the Eddington ratio we adopt the values from Table 4 of \citet{modzelewska2014}, based on \citet{kong2006} formulae and a single component fit: $M = 2.4 \times 10^9 M_{\odot}$, $L_{bol}/L_{Edd} = 0.6$. The object has a typical value of the ${\rm FWHM}$ of $\sim 6000$ km s$^{-1}$, and no clear double symmetric profile, so in order to fit the FRADO model to this data we must adopt higher values of the Planck mean opacity than in the standard model, as discussed in Sect.~\ref{sect:Planck}. We use values of the black hole mass $M=2.4\times10^9M_\odot$ and the accretion rate $\dot m = 0.6$ from \citet{modzelewska2014} and the inclination angle of 45 deg. The results are shown in Fig.~\ref{fig:CTS}.

The main part of the Mg II line is well reproduced by our model, it even well represents the line asymmetry, which is usually modeled as a second independent component. However, the model does not reproduce well the Lorentzian tail on the red wing of the line. Our solution gives the inner radius of the BLR, according to our model, of $6.4 \times 10^{17}$ cm, which is just a little smaller than the $7.6 \times 10^{17}$ cm implied by the \citet{bentz2013} formula, assuming 0.17 dex difference between the logarithm of the monochromatic flux at 3000 \AA~ and 5100 \AA.  

\section{Discussion}

We tested here the FRADO model of the formation of the BLR. The model in its basic form does not have free parameters. Nevertheless, it reproduces well the position of the BLR measured through reverberation method \citep{czerny2016}, and it correctly predicts the typical ratios of the ${\rm FWHM}$ to $\sigma$ characterizing the line profile. It also allows us to calculate the virial factor usually needed to obtain the black hole mass from the line delay and the line width \citep{peterson2004}. We show at the basis of the FRADO model that this factor depends on the black hole mass in the case of enhanced opacity which explains the puzzling dependence of the virial factor on the type of the host galaxy discovered by \citet{hokim2014}. This suggests that the FRADO is a viable mechanism for the formation of the LIL part of the BLR. However, further improvements of the model are needed to ameliorate the current simplified analytical scheme. We hope that those improvements will bring the model closer to the real properties of the BLR.

\subsection{Dust opacity issue}
\label{sect:disc_opac}

The model in its basic form, with a realistic value of the opacity for an optically thin dust, does not lift the material high enough to expose it to the illumination by the central source. This is the most serious problem of the current model. An efficient rise of the material takes place only if we increase artificially the dust opacity in the IR. The question arises whether the dust-driven wind cannot form due to the 'z' component of gravity rising with the distance from the equatorial plane or the model description is too approximate. If the available radiative force acting on dust is indeed not strong enough to create the BLR with large covering factor then other solutions (line driven wind, magnetically driven wind or disk instabilities) must play the dominant role. However, the fact that dust driven wind is efficient in the case of stars, and the location of the BLR based on dust driven wind is appropriate, makes a good motivation for further pursuing the model.

The description of the dynamics used here is purely local, the initial velocity of the material is zero and the driving force must come locally from the underlying accretion disk. Only after the initial rise is the matter is exposed to the radiation from the central source. Thus, if the plasma does not reach higher elevations, it is a serious problem for the description of the opacity. The question thus arises whether the outer disk does not produce enough energy for considerable initial lift of the material, or we do not use it efficiently enough in the current model.

Stellar winds are usually constrained by the amount of momentum carried by radiation.  If this approach is adopted in the case of an accretion disk wind, the maximum allowed local outflow rate is rather low at large distances from the black hole:
\begin{equation}
4 \pi r^2 \dot M_z = {3 \over 2} ({r_g/r})^{1/2} \dot M,
\end{equation}
where $\dot M_z$ is the mass loss from the disk per unit disk surface area. At distances corresponding to the BLR, the amount of material which can be lifted by the radiation momentum is very small. However, if multiple scatterings are allowed and the whole energy can be transferred to the outflow, the situation changes and most of the material can be removed in the form of a wind:
\begin{equation}
4 \pi r^2 \dot M_z = 3 \dot M.
\end{equation} 
Here we assumed that the wind terminal velocity is equal to the local escape velocity which is characteristic for cool stars \citep[e.g.][]{lamers1995}.

This shows that the amount of energy is certainly high enough to push the material up, but it cannot be done using the optically thin approximation for the radiative transfer. Therefore, the approximate treatment of the opacity in our basic model is not at the core of the problem. We need an increase of the radiation pressure by a large factor. 

\begin{figure}
    \centering
    \includegraphics[width=0.95\hsize]{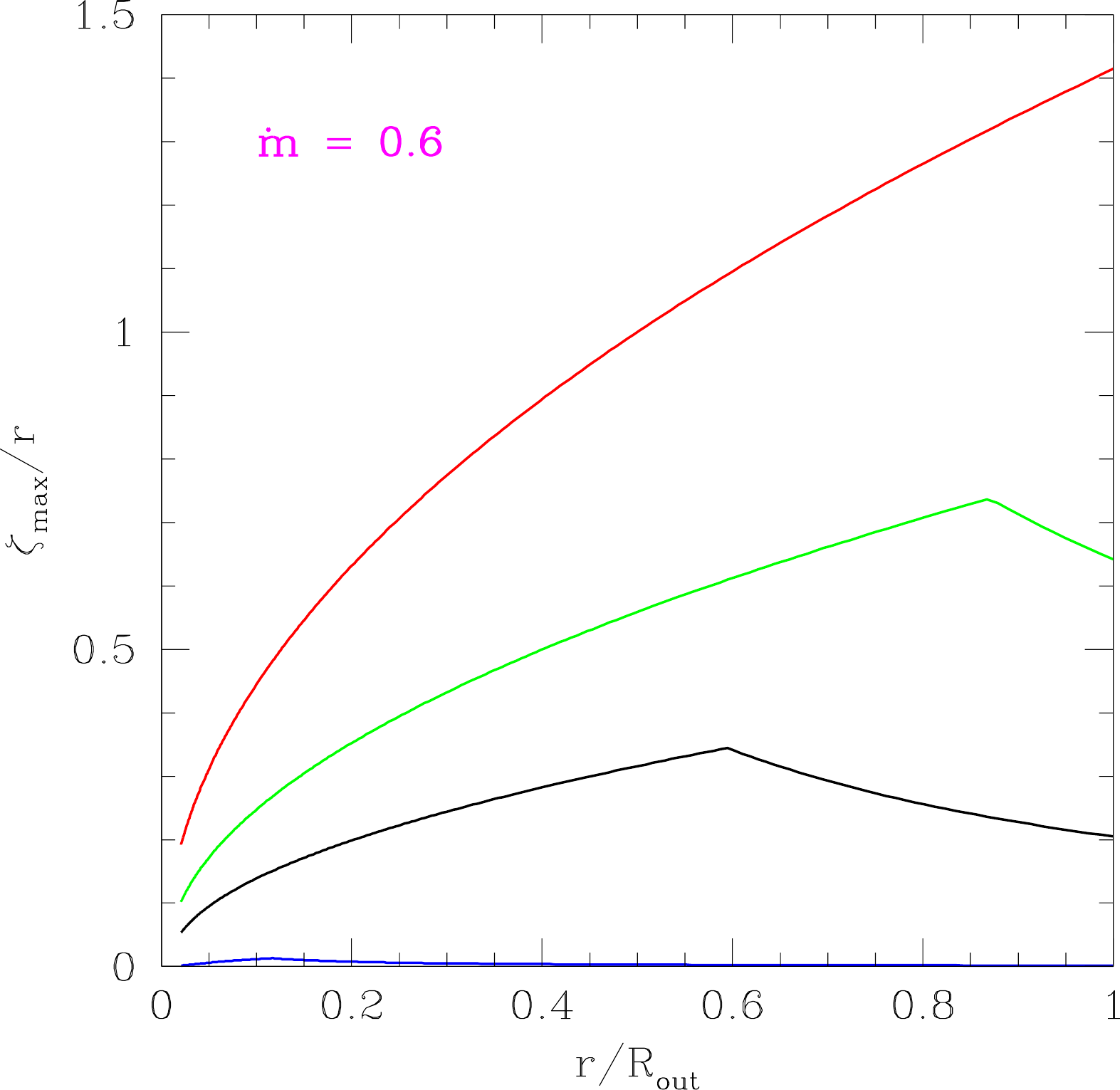}
    \caption{The ratio of the maximum height of the clouds to the current disk radius for the Planck opacity $\kappa_P$ in units cm$^2$g$^{-1}$ equal 8 (blue), 800 (black), 2500 (green) and 8000 (red) for a black hole mass $3 \times 10^7 M_{\odot}$ and accretion rate $\dot m = 0.6$. Small values of opacity (optically thin dust) lead to a small rise of the clouds and an extremely small covering factor, too high values are not consistent with the assumption $z << r$. The outflow and non-local phenomena are most likely at the position where $\zeta_{max}$ reaches the highest value.}
    \label{fig:z_naj}    
\end{figure}

To illustrate this effect we plot the maximum height reached by the clouds as a function of radius for four values of the Planck mean (see Fig.~\ref{fig:z_naj}). This height rises for small radii and finally saturates at constant value, where the dust evaporation does not happen. We see that for the largest value of the Planck mean the saturation does not appear. To reach the height of the radius ratio of 0.3 (and the BLR covering factor of 0.3) we need an opacity of 800 for the adopted black hole mass and the accretion rate. 

This is not simple to achieve. Recent paper discussing the dusty cloud acceleration in the context of starburst galaxies \citep{zhang2017} performed the radiative transfer computations for both optically thin and optically thick clouds, and they find that the acceleration efficiency actually drops with the rise of the optical depth of the cloud. However, wind calculations in AGB stars require much more complex approach, with stellar pulsations playing a key role. The terminal velocities are very low, of order of 50 km s$^{-1}$ \citep[see ][for a grid of wind models]{tashibu2017} although the gravity at the stellar surface is similar to the gravity in AGN at BLR, so the escape velocity is thousands of km s$^{-1}$.

\subsection{Self-shielding in the cloud distribution}

We consider the current description of the individual cloud input to the line profile as the second basic limitation of the analytical model. At present, the cloud height and the shielding of outer clouds by inner clouds is not included. This is an important effect. As we showed in Fig.~\ref{fig:decompo}, lower contribution of the outermost BLR creates much broader emission lines. In this paper we did not attempt to take those effects into account since we believe that we first need an improvement of the description of the radiation pressure acting on dusty clouds as a function of radius. With that at hand, we can model - in the future - the amount of radiation intercepted by the specific cloud self-consistently. This would require supplementing the current description with the cloud size distribution. We could try ad-hoc parametric models or more physically motivated estimates based on thermal instabilities. Current model does not require this information (see Sect.~\ref{sect:profiles}), hence it is indeed very simple.

\subsection{Outflow}

As we already mentioned in Sec.~\ref{sect:disc_opac} (see also Sec.~\ref{sect:Planck}), assumption of an optically thin dust (the opacity given by the Planck mean) leads to a disk-like shape of the emission line due to too small a vertical velocity. By drastically increasing the opacity we can push the material high enough. Then, close to the radial position of the maximum in $\zeta_{max}/r$, our assumptions of small height breaks down. The gravity for $z > r/{\sqrt 2}$ starts to decrease with height and the outflow easily forms. As was shown already by \citet{murray1995} the material does not necessarily escape the system; clouds follow elliptical orbits and fall back towards the disk at a smaller radii. Such a motion cannot be calculated analytically within the frame of our model, but in the future numerical computations of those orbits can be done. Those highest velocity clouds are missing in our analytical model and they will likely provide the Lorentzian wings to the lines. 

\subsection{Disk height and $z/r < 1$ approximation}

In the analysis above we treated the description of the disk thickness in an approximate way and we explicitly assumed that the cloud motion above the disk can be approximated assuming $z/r << 1$. We thus checked if those approximations are correct.

\begin{figure}
    \centering
    \includegraphics[width=0.95\hsize]{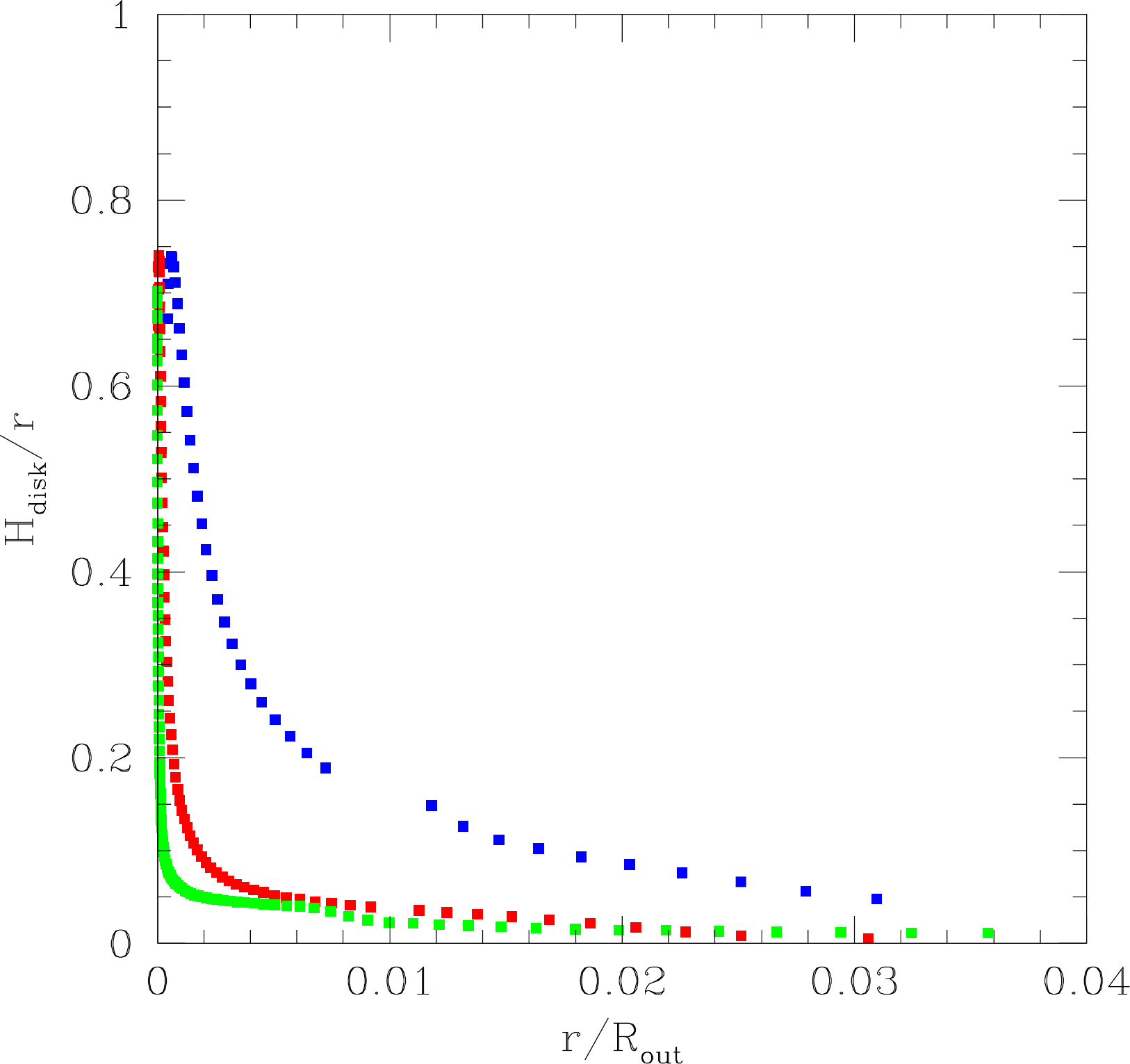}
    \caption{The ratio of the disk thickness to the radius calculated from a realistic model of the disk structure, including the proper description of the opacity and the self-gravity effects for three values of the black hole mass ($3 \times 10^5 M_{\odot}$, the green dots; $3 \times 10^7 M_{\odot}$, the red dots; $3 \times 10^9 M_{\odot}$, the blue dots) and the Eddington accretion rate. The disk is thick only in the innermost region and in the BLR region the disk is geometrically thin.}
    \label{fig:thickness}    
\end{figure}

As for the disk thickness, we used the disk model which describes the vertical structure of a Keplerian disk taking into account the radiative transfer with opacities and the convective vertical transfer as described in \citet{rozanska1999} and with the disk self-gravity effects included (for more details on the model description, see \citet{czerny2016}). We plotted the ratio of the disk thickness to the disk radius as a function of the radius normalized to the outer edge of BLR, $R_{out}$ (see Fig.~\ref{fig:thickness}. We see that the disk itself is geometrically thin in its outer parts. The computations of the disk structure do not cover all the formation region of the BLR, since in that region the effects of self-gravity are too strong to use the standard marginally-self-gravitating disk model. This is a serious problem for the description of the accretion disk there, as was discussed in \citet{czerny2016}, and perhaps the magnetic field plays an important role in this region, or indeed the disk inflow is much faster than in the standard disks and starburst activity plays a dominant role \citep[see e.g.,][]{collin1999,thompson2005,wang2011}. If the disk itself is indeed considerably puffed-up, outflow formation would became easier.

\subsection{Non-local radiation pressure}

In the current picture we neglected the dynamical role of the radiation pressure that comes from other disk radii, in particular coming from the central source. On one hand, this allowed us to get an analytical solution but, on the other hand, this is a serious negligence of the current model. As the clouds rise up, this non-local radiation pressure component becomes increasingly important. If there is no shielding effect, the radiation pressure from the disk central parts dominate at height $\zeta > 1.5 r_g/\eta$ and pushes the material radially. On the other hand, the need for shielding of the outer BLR has been argued by numerous authors in the past \citep{boksenberg1977,murray1995,gallagher2008,gaskell2009,maiolino2010}, and the accretion disk self-shielding is also important in this aspect \citep{czerny1987,wang2014}, which may imply that the assumption used currently in our paper affects the solutions in a less severe way than implied by the constraint above. Accretion disk thickness is a complex function of the radius since various cooling processes operate in the disk, and the proper determination of the shielding would require computing the location of the disk atmosphere \citep[e.g.][]{hryniewicz_2012,rozanska_2014,adhikari2016}. Clearly, the numerical computations of the cloud dynamics are needed, at least at the level represented by QWIND \citep{risaliti2010}, or, preferentially, combining full dynamical computations with a radiative transfer \citep[e.g.][]{namekata2016}.

\subsection{Lorentzian profiles}

The current model usually gives a double peak disk-like profiles.  Single peak skewed profile, or even two-component profiles (see Fig.~\ref{fig:CTS}) can be obtained if we allow for strong cloud acceleration due to the dust pressure. However, we cannot reproduce a single Lorentzian shape yet within our scheme, despite the Lorentzian shape being frequently favored for NLS1s and for a fraction of quasars. It has been argued in the past that Lorentzian shapes are naturally explained by the case of emission from an extended accretion disk \citep{veroncetty2001,sulentic2002}. In our model the emitting region is extended, but nevertheless we do not find broad Lorentzian wings. The Lorentzian shape was modeled in the past by introducing a turbulent motion in outer part of the disk 
\citep{goad2012}. Our model, due to the mixture of inflow and outflow, contains this element, but apparently this is not enough, and we still need locally higher cloud velocities. Perhaps, if we include the real outflow option, this would solve the problem. In the case of supernovae, Lorentzian emission lines are also observed and they are argued to come from optically thick winds with the wings forming in the shocked cold gas and enhanced due to multiple scattering \citep{Chugai2004}. If so, this would imply the proper radiative transfer computations are necessary.

\subsection{Dissipation of the cloud kinetic energy}

The FRADO picture of the BLR formation implies that clouds move both up and down, which leads to collisions between the clouds as well as collisions between the clouds and the disk surface. This effect is ignored in the current model and it would affect both the thermal conditions and the velocity field. The cloud-cloud collisions can give the effect of turbulence, additionally broadening the line profiles. The heat dissipation effect is much more complicated. If the fall-back of the clouds heats the disk surface considerably, dust formation may be temporarily suppressed until the disk cools down, which could result in a strongly time-dependent process, at least in the innermost region of the BLR.   

\subsection{Physics of the stellar dust-driven winds}

The physics of the dust driven winds have been broadly studied in the case of asymptotic giant branch (AGB) stars. The most important aspect of those studies is the conclusion that a stationary wind model cannot explain the observed strong dusty outflow. The key element of any such stellar wind model is the inclusion of the stellar pulsations. Those pulsations temporarily lift the material up to the lower gravity region, cool it and lead also to creation of strong shocks. The dust condensation takes place at a considerable height above the stellar surface and only there can the radiation pressure lead to a massive outflow \citep[e.g.][]{schirrmacher2003,liljegren2016}.

The time-dependent aspect can be important for AGN outflows as well. As already shown by \citet{watersproga2016} and \citet{freytag2017} for a line-driven wind, strong variability of the AGN continuum helps to increase the cloud acceleration efficiency. However, it requires the variability to happen on timescales longer than the local thermal timescale. Whether the conditions in the outer parts of the disk are likely to satisfy those conditions should be studied in the future.

\subsection{Alternative mechanisms for rising up BLR material}

If the FRADO model in the more advanced formulation fails to produce the BLR with a large covering factor caused by the material present high above the disk, then it will be indeed necessary to apply supporting mechanisms or alternatives. The FRADO model neglects other mechanisms, which can support or even replace the radiation pressure force acting on dust. The HIL part formation of the BLR is clearly driven by line pressure, and some line pressure can work also in the LIL part. The disk atmosphere is also likely magnetized, the dominant role of the magnetic field has been discussed in a number of BLR and dusty torus models \citep[e.g.][]{dorodnitsyn2017}. Another aspect is the interaction of the stars and the accretion disk - stars passing through the disk can lead to disk disturbances \citep[e.g.][]{vilkoviskij2002}, with disk material tracing the stars and possibly giving the effect of the BLR \cite{zurek1994,karas2001}. Also, it has been argued that the effect of
dust charging helps to levitate grains above the equatorial plane, thus counter-acting the collapsing role of
self-gravity in magnetized disks and tori \citep{kovar2011,trova2016}. 
Finally, the role of self-gravity in the outer disk is not clear \citep{laor1989,collin1999}, and it might also contribute to pushing some material out of the equatorial plane \citep{wang2011,hopkins2012,czerny2016}. The processes happening in the outer part of an accretion disk are thus complex, but we think that the FRADO model catches the most important properties of the LIL BLR and forms an interesting starting point for the future studies.  

\section{Conclusions}

The FRADO model based on radiation pressure acting on the dusty disk atmosphere has considerable predictive power. The model depends only on the global parameters of an active nucleus and the viewing angle. The model
\begin{itemize}
\item uniquely predicts the inner and the outer radius of the BLR
\item gives the line profile asymmetry due to the dust evaporation in the failed wind  
\item gives the line shape parameter $FWHM/\sigma$ consistent with the data
\item predicts the dependence of the virial factor on the black hole mass.
\end{itemize} 
Further developments are still necessary in the description of the radiation pressure to reproduce well the BLR covering factor, and to provide more realistic line shapes.

\acknowledgments
{\small
We are grateful to the referee for the comments which helped to clarify the paper. The project was partially supported by the Polish Funding Agency National Science Centre, project 2015/17/B/ST9/03436/ (OPUS 9). VK acknowledges the Czech Ministry of Education, Youth and Sports project LD15061 ``Astrophysics of toroidal fluid structures''. Y.R.L. acknowledges financial support from the National Natural Science 
Foundation of China (NSFC; 11570326) and from the Strategic Priority 
Research Program of the Chinese Academy of Sciences (XDB23000000). 
J.M.W. acknowledges financial support from the Ministry of Science and 
Technology of China (2016YFA0400700), from the Key Research Program of 
Frontier Sciences, CAS (QYZDJ-SSW-SLH007), and from NSFC grants 
(11233003 and U1431228). The project also received funding from the European Union Seventh Framework Program (FP7/2007-2013) under the grant agreement No.312789. Based  on  observations  made  with  the  Southern  African  Large Telescope (SALT) under program 2012-2-POL-003, 2013-1-POL-RSA-002, 2013-2-POL-RSA-001, 2014-1-POL-RSA-001, 2014-2-SCI-004, 2015-1-SCI-006, 2015-2-SCI-017, 2016-1-SCI-011, and 2016-2-SCI-024
 (PI: B. Czerny).
}
\bibliographystyle{aasjournal}
\bibliography{dust}
\end{document}